\documentclass[12pt,final]{l4dc2021} 

 \newcommand{\zono}[1]{\langle #1 \rangle}
\DeclareMathOperator*{\argminB}{argmin}   
\DeclareMathOperator*{\argmaxB}{argmax} 
\usepackage{times}
\usepackage{tikz}
\usepackage{graphicx}
\newtheorem{assumption}{Assumption} 

\newtheorem{theorem2}{Theorem} 
\newtheorem{definition2}{Definition} 
\newtheorem{remark2}{Remark} 

\makeatletter
\def\set@curr@file#1{\def\@curr@file{#1}} 
\makeatother
\usepackage[load-configurations=version-1]{siunitx} 
\usepackage{epstopdf}

\title[Data-Driven Reachability]{Data-Driven Reachability Analysis Using Matrix Zonotopes}

\author{%
 \Name{Amr Alanwar} \Email{alanwar@kth.se}\\
   \addr{KTH Royal Institute of Technology}\\
 \Name{Anne Koch} \Email{anne.koch@ist.uni-stuttgart.de}\\
 \addr{University of Stuttgart}\\
 \Name{Frank Allgöwer} \Email{frank.allgower@ist.uni-stuttgart.de}\\
 \addr{University of Stuttgart}\\
 \Name{Karl Henrik Johansson} \Email{kallej@kth.se}\\
  \addr{KTH Royal Institute of Technology} \vspace{-6mm}
}

\begin{document}
\maketitle

\begin{abstract}
In this paper, we propose a data-driven reachability analysis approach for unknown system dynamics. Reachability analysis is an essential tool for guaranteeing safety properties. However, most current reachability analysis heavily relies on the existence of a suitable system model, which is often not directly available in practice. We instead propose a data-driven reachability analysis approach from noisy data. More specifically, we first provide an algorithm for over-approximating the reachable set of a linear time-invariant system using matrix zonotopes. Then we introduce an extension for Lipschitz nonlinear systems. We provide theoretical guarantees in both cases. Numerical examples show the potential and applicability of the introduced methods. 
\end{abstract}
\begin{keywords}%
  Reachability analysis, data-driven methods, zonotope. %
\end{keywords}

\section{Introduction}

Reachability analysis computes the reachable set, which is the union of all possible trajectories that a system can reach within a finite or infinite time when starting from a bounded set of initial states, subject to a set of possible inputs \citep{conf:thesisalthoff}. Most of the existing reachability analysis techniques assume the availability of a model. However, systems are becoming more complex, and data is becoming more readily available. Therefore, we consider the problem of computing reachable sets directly from noisy data without the need for a model. 


\begin{figure}[h]
\begin{center}
\includegraphics[scale=0.25]{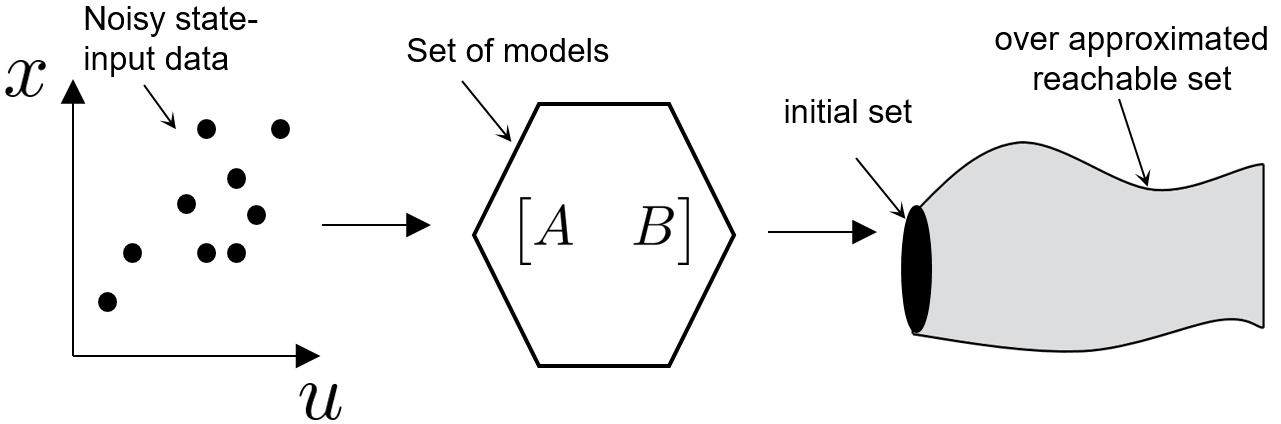}
\end{center}
\vspace{-4mm}
    \caption{We compute the reachable set consistent with noisy input-state data.}
\label{fig:idea}
\vspace{-3mm}
\end{figure}

The most popular approaches in computing reachable sets are
set-propagation and simulation-based techniques. 
Set-propagation techniques propagate reachable sets for consecutive time points. The efficiency of these methods depends on the set representation and the computational technique. 
Popular set representations are polyhedra \citep{conf:reachpolydt,conf:reachpoly1}, zonotopes \citep{conf:thesisalthoff,conf:reachabilityzono}, (sparse) polynomial zonotopes \citep{conf:reachpolyzono,conf:sparsepolyzono}, ellipsoids \citep{conf:reachellipsoidal}, support functions \citep{conf:reachsupport,conf:reachsupportlinear}. Zonotopes have favorable properties as they can be represented compactly, and they are closed under the Minkowski sum and linear mapping. The simulation-based approach in \citep{conf:reachsim3} over-approximates the reachable set by a collection of simulation tubes around trajectories, such that the union of these tubes provides an over-approximation of the reachable set. Other simulation-based techniques are proposed in \citep{conf:reachsim2,conf:reachsimparsimonious,conf:reachsim4,conf:reachsim1,conf:reachsimmatrixmeasures,conf:reachrandomsettheory}. 
Another approach in \citep{conf:murat} finds an under-approximation for the reachable set from data with an unknown system model. However, their approach is only \emph{probabilistically} accurate, i.e., the more data is sampled, the higher is the probability that the reachable set is correct. While there hence exist efficient reachability algorithms for a given model and some initial approaches for reachability analysis directly from data, obtaining a guaranteed over-approximation of the reachable set from noisy data of an unknown model is still an open problem.

With the rising amount of available data, the interest in data-driven methods for analysis and control has increased, see for example \citep{conf:deepc,conf:nonexciting,conf:nonlinear1,conf:switched,conf:mpcguarantees}. In this work, we will specifically make use of ideas used in \citep{conf:nonexciting,conf:dissipativity1,conf:dissipativity2} and \citep{conf:formulas,conf:annerobustcontrol,conf:berberich2020combining,conf:controlslemma} for data-driven analysis and data-driven controller design, respectively. In these works, the data is generally used to provide a characterization of all models that are consistent with the data. This characterization is chosen in a way to provide a computational approach for direct systems analysis and design without explicitly identifying a model. 

Using efficient computational tools from model-based reachability analysis together with recent development in data-driven systems analysis and control, we propose in this paper a technique to compute a guaranteed reachable set directly from noisy data as visualized in Figure \ref{fig:idea}. Providing such guarantees from noisy data requires the computation of the set, which encloses all models that are consistent with the noisy data. To this end, we use matrix zonotopes since they can be used in a computationally efficient way for forward propagation as they are closed under Minkowsky sum and linear mapping. All used codes to recreate our findings are publicly available\footnotemark. 
The contributions of this paper can be summarized as follows:

\footnotetext{\href{https://github.com/aalanwar/Data-Driven-Reachability-Analysis}{https://github.com/aalanwar/Data-Driven-Reachability-Analysis}}

\begin{itemize}
    \item We provide a method using matrix zonotopes to over-approximate the reachable set of an unknown linear control system from noise-corrupted input-state data (Theorem \ref{th:reach_lin}). 
    \item The method is extended to over-approximate the reachable set of Lipschitz nonlinear systems (Theorem \ref{th:reachdisnonlin}). 
\end{itemize}

The rest of the paper is organized as follows: the preliminaries and problem statement are introduced in Section~\ref{sec:pb}. Data-driven reachability analysis for linear systems is proposed in  Section~\ref{sec:reachlineardis}. Then, we extend the proposed approach to Lipschitz nonlinear systems in Section~\ref{sec:reachnonlinear}. The introduced approaches are applied to numerical examples in Section~\ref{sec:eval} and Section~\ref{sec:con} concludes the paper.

\section{Preliminaries and Problem Statement} \label{sec:pb}





We start by defining our set representations used in the reachability analysis. 

\begin{definition2}(\textbf{Zonotope} \citep{conf:zono1998}) \label{def:zonotopes} 
Given a center $c_\mathcal{Z} \in \mathbb{R}^n$ and $\gamma_\mathcal{Z} \in \mathbb{N}$ generator vectors in a generator matrix $G_\mathcal{Z}=[g_\mathcal{Z}^{(1)},...,g_\mathcal{Z}^{(\gamma_\mathcal{Z})}] \in \mathbb{R}^{n \times \gamma_\mathcal{Z}}$, a zonotope is defined as
\begin{equation}
	\mathcal{Z} = \Big\{ x \in \mathbb{R}^n \; \Big| \; x = c_\mathcal{Z} + \sum_{i=1}^{\gamma_\mathcal{Z}} \beta^{(i)} \, g_\mathcal{Z}^{(i)} \, ,
	-1 \leq \beta^{(i)} \leq 1 \Big\} \; .
\end{equation}
We use the shorthand notation $\mathcal{Z} = \zono{c_\mathcal{Z},G_\mathcal{Z}}$. 
\end{definition2}

A linear map $L$ is defined as $L \mathcal{Z}  = \zono{L c_\mathcal{Z}, L G_\mathcal{Z}}$. Given two zonotopes $\mathcal{Z}_1=\langle c_{\mathcal{Z}_1},G_{\mathcal{Z}_1} \rangle$ and $\mathcal{Z}_2=\langle c_{\mathcal{Z}_2},G_{\mathcal{Z}_2} \rangle$, the Minkowski sum is: 
$\mathcal{Z}_1 + \mathcal{Z}_2 = \Big\langle c_{\mathcal{Z}_1} + c_{\mathcal{Z}_2}, \begin{bmatrix} G_{\mathcal{Z}_1} & G_{\mathcal{Z}_2} \end{bmatrix}\Big\rangle.$
For simplicity, we use the notation $+$ instead of $\oplus$ for Minkowski sum as the type can be determined from the context. Similarly, we use  $\mathcal{Z}_1 - \mathcal{Z}_2$ to denote $\mathcal{Z}_1 + -1 \mathcal{Z}_2$. We define the Cartesian product of two zonotopes $\mathcal{Z}_1 $ and $\mathcal{Z}_2$ by 
\begin{align}\label{eq:cart}
\mathcal{Z}_1 \times \mathcal{Z}_2 = \bigg\{ \begin{bmatrix}z_1 \\ z_2\end{bmatrix} \bigg| z_1 \in \mathcal{Z}_1, z_2 \in \mathcal{Z}_2 \bigg\} = \Big\langle \begin{bmatrix} c_{\mathcal{Z}_1} \\ c_{\mathcal{Z}_2} \end{bmatrix}, \begin{bmatrix} G_{\mathcal{Z}_1} & 0 \\ 0 & G_{\mathcal{Z}_2} \end{bmatrix} \Big\rangle.
\end{align}

\begin{definition2}\label{def:matzonotopes}(\textbf{Matrix Zonotpe} \citep[p.52]{conf:thesisalthoff})  
Given a center matrix $C_{\mathcal{M}} \in \mathbb{R}^{n\times T}$ and $\gamma_\mathcal{M} \in \mathbb{N}$ generator matrices $\tilde{G}_{\mathcal{M}}=[G_{\mathcal{M}}^{(1)},\dots,G_{\mathcal{M}}^{(\gamma_\mathcal{M})}] \in \mathbb{R}^{n \times (T \times \gamma_\mathcal{M})}$, a matrix zonotope is defined as
\begin{equation}
	\mathcal{M} = \Big\{ X \in \mathbb{R}^{n\times T} \; \Big| \; X = C_{\mathcal{M}} + \sum_{i=1}^{\gamma_\mathcal{M}} \beta^{(i)} \, G_{\mathcal{M}}^{(i)} \, ,
	-1 \leq \beta^{(i)} \leq 1 \Big\} \; .
\end{equation}
We use the shorthand notation $\mathcal{M} = \zono{C_{\mathcal{M}},\tilde{G}_{\mathcal{M}}}$. 
\end{definition2}

\begin{definition2}\label{def:intmat}(\textbf{Interval Matrix} \citep[p. 42]{conf:thesisalthoff})
An interval matrix $\mathcal{I}$ specifies the interval of all possible values for each matrix element between the left limit $\underline{I}$ and right limit $\bar{I}$:
\begin{align}
    \mathcal{I} = \begin{bmatrix} \underline{I},\bar{I}  \end{bmatrix}, \quad \b{I},\bar{I} \in \mathbb{R}^{n\times n}
\end{align}
\end{definition2}

The conversion of a matrix zonotope $\mathcal{M}$ to an interval matrix is denoted by $\mathcal{I} = \begin{bmatrix} \underline{I} , \bar{I}  \end{bmatrix}$ by writing: $\begin{bmatrix} \underline{I} , \bar{I}  \end{bmatrix} =\text{intervalMatrix}(\mathcal{M})$. Similarly, we write $\text{zonotope}$ to convert an interval to a zonotope. 

Consider a discrete-time linear system
\begin{equation}
    \begin{split}
        x(k+1) &= A x(k) + B u(k)+  w(k),
    \end{split}
    \label{eq:sys}
\end{equation}
where $A \in \mathbb{R}^{n \times n}$ and $B \in \mathbb{R}^{n \times m}$ are the system dynamics, 
$w(k) \in \mathcal{Z}_w = \zono{c_{\mathcal{Z}_w},G_{\mathcal{Z}_w}}\subset \mathbb{R}^n$ denotes the bounded noise by a noise zonotope $\mathcal{Z}_w$, $u(k) \in \mathcal{U}_k \subset \mathbb{R}^{m}$ the input bounded by an input zonotope $\mathcal{U}_k$, and $x(0) \in \mathcal{X}_0 \subset \mathbb{R}^n$ the initial state of the system bounded by the initial set $\mathcal{X}_0$. 
We aim to compute the reachable set when the model of the system in \eqref{eq:sys} is unknown, but input and noisy state trajectories are available. More specifically, we consider $K$ input-state trajectories of different lengths $T_i$, $i=1,\dots,K$, denoted by $\{u^{(i)}(k)\}_{k=0}^{T_i - 1}$, $\{x^{(i)}(k)\}_{k=0}^{T_i}$, $i=1, \dots, K$. We collect the set of all data sequences in the following matrices
 \begin{align*}
     X &=  \begin{bmatrix} x^{(1)}(0) \; \dots \; x^{(1)}(T_1) & 
     x^{(2)} (0) \; \dots \; x^{(2)}(T_2) \; \dots \; x^{(K)}(0) \; \dots \; x^{(K)}(T_K)\end{bmatrix}, \\
     U_- &= \begin{bmatrix} u^{(1)}(0) \; \dots \; u^{(1)}(T_1-1) &
     u^{(2)} (0) \; \dots \; u^{(2)}(T_2-1) \; \dots \; u^{(K)}(0) \; \dots \; u^{(K)}(T_K-1) \end{bmatrix}.
 \end{align*}
 Let us further denote
 \begin{align*}
     X_+ &= \begin{bmatrix} x^{(1)}(1) \; \dots \; x^{(1)}(T_1) & 
     x^{(2)} (1) \; \dots \; x^{(2)}(T_2) \; \dots \; x^{(K)}(1) \; \dots \; x^{(K)}(T_K) \end{bmatrix}, \nonumber\\
     X_- &= \begin{bmatrix} x^{(1)}(0) \; \dots \; x^{(1)}(T_1{-}1) & x^{(2)}(0) \; \dots \; x^{(2)}(T_2{-}1) \; \dots \; x^{(K)}(0) \; \dots \; x^{(K)}(T_K{-}1) \end{bmatrix}.
 \end{align*}
The total amount of data points from all available trajectories is denoted by $T = \sum_{i=1}^{K} T_i$ and we denote the set of all available data by $\mathcal{D}=\{U_-, X\}$. 
Reachability analysis computes the set of states $x$ which can be reached given a set of uncertain initial states $\mathcal{X}_0 \subset \mathbb{R}^n$
containing the initial state $x(0) \in \mathcal{X}_0$ and a set of uncertain inputs
$\mathcal{U}_k \subset \mathbb{R}^m$ containing the inputs $u(k) \in \mathcal{U}_k $.

\begin{definition2} 
The reachable set $\mathcal{R}_{k}$ after $N$ time steps, inputs $u(k) \in \mathcal{U}_k  \subset \mathbb{R}^m, \forall k \in \{0,...,N{-}1\}$, noise $w( \cdot) \in \mathcal{Z}_w$, and initial set $\mathcal{X}_0 \in  \mathbb{R}^n$, is the set of all states trajectories starting in $\mathcal{X}_0$ after $N$ steps: 
\begin{align} \label{eq:R}
        \mathcal{R}_{N} = \big\{ x(N) \in \mathbb{R}^n \, \big|& x(k{+}1) = Ax(k) {+} Bu(k) + w(k), 
        \, x(0) \in \mathcal{X}_0,\nonumber\\
        &u(k) \in \mathcal{U}_k, w(k) \in \mathcal{Z}_w: \forall k \in \{0,...,N{-}1\}\big\}.
\end{align}
\end{definition2}



Let us denote the actual noise in the data by $\hat{w}$. From the bounded noise assumption, it follows directly that the stacked matrix
\begin{align*}
    \hat{W}_- = \begin{bmatrix} 
    \hat{w}^{(1)}(0) \; \dots \; \hat{w}^{(1)}(T_1{-}1) & \hat{w}^{(2)}(0) \; \dots \; \hat{w}^{(2)}(T_2{-}1) \; \dots \; \hat{w}^{(K)}(0) \; \dots \; \hat{w}^{(K)}(T_K{-}1)\end{bmatrix}
\end{align*}
is an element of the set $\hat{W}_- \in \mathcal{M}_w$ where $\mathcal{M}_w =\zono{ C_{\mathcal{M}_w}, [G_{\mathcal{M}_w}^{(1)},\dots,G_{\mathcal{M}_w}^{(\gamma_{\mathcal{Z}_w} T)}] }$ is the matrix zonotope resulting from the concatenation of multiple noise zonotopes $\mathcal{Z}_w=\zono{c_{\mathcal{Z}_w},\begin{matrix} g_{\mathcal{Z}_w}^{(1)} &\dots& g_{\mathcal{Z}_w}^{(\gamma_{\mathcal{Z}_w})}\end{matrix}}$ as follows:
\begin{align*}
    C_{\mathcal{M}_w} &= \begin{bmatrix}c_{\mathcal{Z}_w} & \dots & c_{\mathcal{Z}_w}\end{bmatrix}, \quad
    &&G^{(1+(i-1)T)}_{\mathcal{M}_w} = \begin{bmatrix} g_{\mathcal{Z}_w}^{(i)} & 0_{n \times  (T-1)}\end{bmatrix}, \\
    G^{(j+(i-1)T)}_{\mathcal{M}_w} &= \begin{bmatrix} 0_{n \times  (j-1)} &g_{\mathcal{Z}_w}^{(i)}  & 0_{n \times  (T-j)}\end{bmatrix}, \quad
    &&G^{(T+(i-1)T)}_{\mathcal{M}_w} = \begin{bmatrix} 0_{n \times (T-1)}& g_{\mathcal{Z}_w}^{(i)}\end{bmatrix}.
\end{align*}
$\forall i =\{1, \dots, \gamma_{\mathcal{Z}_w}\}$, $j=\{2,\dots,T-1\}$. We denote the Kronecker product by $\otimes$. We also denote the element at row $i$ and column $j$ of matrix $A$ by $(A)_{i,j}$ and column $j$ of $A$ by $(A)_{.,j}$. For vectors, we denote the element $i$ of vector $a$ by $(a)_i$. We define also for $N$ time steps
\begin{align}
    \mathcal{F} = \cup_{k=0}^{N} (\mathcal{R}_k \times \mathcal{U}_k).
    \label{eq:F}
\end{align}
Finally, we denote all system matrices $\begin{bmatrix} A & B \end{bmatrix}$ that are consistent with the data $\mathcal{D} = (U_-,X)$ by $\mathcal{M}_{{A,B}}$:
\begin{align*}
    \mathcal{M}_{{A,B}} = \{ \begin{bmatrix} A & B \end{bmatrix} | \; X_+ = A X_- + B U_- + W_-, \quad W_- \in \mathcal{M}_w \}.
\end{align*}




\section{Reachability Analysis for Linear Systems}\label{sec:reachlineardis}

Due to the presence of noise, there generally exist multiple matrices $\begin{bmatrix}A & B \end{bmatrix}$ that are consistent with the data. 
To provide reachability analysis guarantees, we need to consider all models that are consistent with the data. Therefore, we are interested in computing a set $\mathcal{M}_\Sigma$ that contains all possible $\begin{bmatrix}A & B \end{bmatrix}$ that are consistent with the input-state measurements and the given noise bound. We apply ideas from \citep{conf:dissipativity1} to our zonotopic noise descriptions, which yields a \textit{matrix zonotope} $\mathcal{M}_\Sigma \supseteq \mathcal{M}_{A,B}$ 
paving the way to computationally simple reachability analysis.

%
%
\begin{theorem2}
\label{th:reachdis}
Given input-state trajectories $\mathcal{D} = (U_-,X)$ of the system \eqref{eq:sys} and a matrix $H$ such that 
\begin{align}
    \begin{bmatrix} 
    X_- \\ U_- 
    \end{bmatrix} H = I,
    \label{eq:cond}
\end{align}
then the matrix zonotope 
\begin{align}
    \mathcal{M}_\Sigma = (X_{+} - \mathcal{M}_w) H
    \label{eq:zonoAB}
\end{align} 
 contains all matrices $\begin{bmatrix}A & B \end{bmatrix}$ that are consistent with the data $\mathcal{D} = (U_-,X)$ and the noise bound, i.e., $\mathcal{M}_{A,B} \subseteq \mathcal{M}_\Sigma$. 
\end{theorem2}
\begin{proof}
For any $\begin{bmatrix}A & B \end{bmatrix} \in \mathcal{M}_\Sigma$, 
we know that there exists a $W_- \in \mathcal{M}_w$ such that 
\begin{align}
    A X_- + B U_- = X_+ - W_-. 
    \label{eq:pf1}
\end{align}
Every $W_- \in \mathcal{M}_w$ can be represented by a specific choice of $\beta^{(i)}_{W_-}$, $-1 \leq \beta^{(i)}_{W_-} \leq 1$, $i=1,\dots,\gamma_\mathcal{Z} T$, that results in a matrix inside the matrix zonotope $\mathcal{M}_w$:
\begin{align}
    W_- &= C_{\mathcal{M}_w} + \sum_{i=1}^{\gamma_\mathcal{Z} T} \beta^{(i)}_{W_-} G_{\mathcal{M}_w}^{(i)}.
    \label{eq:wzono}
\end{align}
Multiplying $H$ from the right to both sides in \eqref{eq:pf1} yields
\begin{align}
   \begin{bmatrix} A & B \end{bmatrix} = \left( X_{+} -  C_{\mathcal{M}_w} +\sum_{i=1}^{\gamma_\mathcal{Z} T} \beta^{(i)}_{W_-} G_{\mathcal{M}_w}^{(i)} \right) H. 
   \label{eq:pf2}
\end{align}
Hence, for all $\begin{bmatrix}A & B \end{bmatrix} \in \mathcal{M}_{A,B}$, 
there exists $\beta^{(i)}_{W_-}$, $-1 \leq \beta^{(i)}_{W_-} \leq 1$, $i=1,\dots,T$, such that \eqref{eq:pf2} holds and hence all $\begin{bmatrix} A & B \end{bmatrix} \in \mathcal{M}_\Sigma$ as defined in \eqref{eq:zonoAB}, which concludes the proof.
%
%
%
\end{proof}

\begin{remark2}
Condition \eqref{eq:cond} in Theorem \ref{th:reachdis} requires that there exists a right-inverse of the matrix $\begin{bmatrix} X_- \\ U_- \end{bmatrix}$. This is equivalent to requiring this matrix to have full row rank, i.e., $\mathrm{rank} \begin{bmatrix} X_- \\ U_- \end{bmatrix} = n+m$, which can be easily checked given the available data $\mathcal{D}$. Note that for noise-free measurements, this rank condition can also be enforced by requiring \eqref{eq:sys} to be controllable and choosing the input persistently exciting of order $n+1$ (compare to \cite{conf:willems}).
\end{remark2}

To guarantee an over-approximate reachable set for the unknown system, we need to consider the union of reachable sets of all $\begin{bmatrix}A & B \end{bmatrix}$ that are consistent with the data. We apply the results of Theorem~\ref{th:reachdis} and do reachability analysis to all systems in the set $\mathcal{M}_\Sigma$. Let $\mathcal{R}_{k}$ denote the reachable set computed based on the true model and $\hat{\mathcal{R}}_{k}$ the reachable set computed based on the noisy data. We can compute $\hat{\mathcal{R}}_{k}$ as an over-approximation of $\mathcal{R}_{k}$ as follows:
\begin{theorem2}
\label{th:reach_lin}
Given input-state trajectories $\mathcal{D} = (U_-,X)$ of the system in \eqref{eq:sys} and a matrix $H$ as defined in~\eqref{eq:cond}, then 
\begin{align}
\hat{\mathcal{R}}_{k+1} = \mathcal{M}_{\Sigma} (\hat{\mathcal{R}}_{k} \times \mathcal{U}_{k}  ) +  \mathcal{Z}_w, \quad \hat{\mathcal{R}}_{0}=\mathcal{X}_0
\end{align}
contains the model-based reachable set, i.e., $\hat{\mathcal{R}}_{k} \supset \mathcal{R}_{k}$. 
\end{theorem2}

\begin{proof} The reachable set computed based on the model can be found using
\begin{align}
\mathcal{R}_{k+1} &=\begin{bmatrix} A & B \end{bmatrix} (\mathcal{R}_{k} \times \mathcal{U}_{k} )+  \mathcal{Z}_w.
\end{align}
Since $\begin{bmatrix} A & B \end{bmatrix} \in \mathcal{M}_{\Sigma}$ according to Theorem \ref{th:reachdis} and both $\mathcal{R}_k$ and $\hat{\mathcal{R}}_k$ start from the same initial set $\mathcal{X}_0$, it holds that $\mathcal{R}_{k+1} \subset \hat{\mathcal{R}}_{k+1}$.
\end{proof}

\section{Reachability Analysis for Lipschitz Nonlinear Systems}\label{sec:reachnonlinear}



We consider a Lipschitz nonlinear system
\begin{align}
    x(k+1) &= f(x(k),u(k))+ w(k),
    \label{eq:sysnonlin}
\end{align}
where we assume $f$ to be twice differentiable. A local linearization of \eqref{eq:sysnonlin} is performed by a Taylor series expansion around the linearization point $z^\star=\begin{bmatrix}x^\star \\u^\star \end{bmatrix}$:
\begin{align}
f(z) = f(z^\star) + \frac{\partial f(z)}{\partial z}\Big|_{z=z^\star} (z - z^\star) + \frac{1}{2}(z - z^\star)^T \frac{\partial^2 f(z)}{\partial z^2}\Big|_{z=z^\star}(z - z^\star) + \dots
\end{align}
The infinite Taylor series can be over-approximated by a first-order Taylor series and a remainder term $L(z)$  \citep{conf:taylor} with

\begin{align}
f(z) \in f(z^\star) + \frac{\partial f(z)}{\partial z}\Big|_{z=z^\star} (z - z^\star) 
+ L(z).
\label{eq:linfL}
\end{align}
In model-based approaches, the term $L(z)$ is usually bounded by the Lagrange remainder \citep{conf:thesisalthoff}
\begin{align*}
    L(z) = \frac{1}{2}(z - z^\star)^T \frac{\partial^2 f(\zeta)}{\partial z^2} (z - z^\star)
\quad \text{with} \quad
\zeta \in \{ z^\star + \alpha ( z-z^\star) | \alpha \in [0,1] \}.
\end{align*}
Since the model is assumed to be unknown, we aim to over-approximate $L(z)$ from data. We rewrite \eqref{eq:linfL} as follows:

\begin{align}
f(x,u) = f(x^\star,u^\star) + \underbrace{\frac{\partial f(x,u)}{\partial x}\Big|_{x=x^\star,u=u^\star}}_{\tilde{A}} (x - x^\star) + \underbrace{\frac{\partial f(x,u)}{\partial u}\Big|_{x=x^\star,u=u^\star}}_{\tilde{B}} (u - u^\star) + L(x,u)\nonumber,
\end{align}
i.e.,
\begin{align}
f(x,u) = \begin{bmatrix}f(x^\star,u^\star) & \tilde{A} & \tilde{B}\end{bmatrix} \begin{bmatrix}1\\x-x^\star\\ u-u^\star\end{bmatrix} +L(x,u).
\label{eq:fz_incl}
\end{align}

If a model of the system is available, the Lagrange remainder $L(z)$ can be over-approximated by an interval which can be converted to a zonotope  \citep{conf:thesisalthoff}. In the following, we apply similar idea from a data-driven viewpoint. More specifically, we conduct data-driven reachability analysis for nonlinear systems by the following two steps: 
\begin{enumerate}
    \item Obtain an approximate linearized model from the noisy data.
    \item Obtain a zonotope that over-approximates the modeling mismatch together with the Lagrange remainder $L(z)$ for the chosen system. 
\end{enumerate}

To obtain an approximate linearized model, we apply a least-squares approach. Without additional knowledge on $L(z)$ and $\hat{w}(k) \in \mathcal{Z}_w$ (or $\hat{W}_- \in \mathcal{M}_w = \langle C_{\mathcal{M}_w}, \tilde{G}_{\mathcal{M}_w}  \rangle$), a best guess in terms of a least-square approach is
\begin{align}
    \tilde{M} = (X_+ - C_{\mathcal{M}_w}) D
    \label{eq:M_tilde}
\end{align}
 where 
 \begin{align}
    \begin{bmatrix} 
    1_{1 \times T}\\ X_{-}-1 \otimes x^\star \\ U_{-}-1 \otimes u^\star 
    \end{bmatrix} D = I,
    \label{eq:cond2}
\end{align}
with the assumption that the right-inverse $D$ exists.


To over-approximate the remainder term $L(z)$ from data, we need to assume that $f$ is Lipschitz continuous for all $z$ in the reachable set $\mathcal{F}$ as defined in \eqref{eq:F}.

\begin{assumption}
It holds that $f: \mathcal{F} \rightarrow \mathbb{R}^n$ is Lipschitz continuous, i.e., that there is some $L^\star \geq 0$ such that 
$\| f(z) - f(z^{\prime}) \|_2 \leq L^\star \| z - z^{\prime}\|_2$
holds for all $z, z^{\prime} \in \mathcal{F}$.
\label{as:lipschitz}
\end{assumption}
For data-driven methods of nonlinear systems, Lipschitz continuity is a common assumption (e.g. \cite{conf:montenbruckLipschitz,conf:novaraLipschitz}). 
%
%
By compactness of $\mathcal{U}_k $, $\mathcal{R}_k$, $k=0,\dots,N$, also $\mathcal{F}$ is compact. Therefore, the data points $\mathcal{D} = (U_-,X)$ are relatively dense in $\mathcal{F}$ such that for any $z \in \mathcal{F}$  there exists a $z_i = \begin{bmatrix} (X_-)_{\cdot,i} \\ (U_-)_{\cdot,i} \end{bmatrix} \in \mathcal{D}$ such that $\|z - z_i \| \leq \delta$. The quantity $\delta$ is sometimes referred to as the covering radius or the dispersion. The following theorem over-approximates the reachable sets of \eqref{eq:sysnonlin}.




\begin{theorem2}
\label{th:reachdisnonlin}
Given data $\mathcal{D} = (U_-,X)$, an over-approximation of the model-based reachable set $\hat{\mathcal{R}}_{k} \supset \mathcal{R}_{k}$ of \eqref{eq:sysnonlin} starting from $\hat{\mathcal{R}}_{k}=\mathcal{X}_0$ can be computed as follows
\begin{align}
\hat{\mathcal{R}}_{k+1} = \tilde{M} (1 \times \hat{\mathcal{R}}_{k} \times \mathcal{U}_k ) +  \mathcal{Z}_w +  \mathcal{Z}_L + \mathcal{Z}_\epsilon,
\end{align}
with $\tilde{M}$ as defined in \eqref{eq:M_tilde}, and
\begin{align}
\mathcal{Z}_L  = \textup{zonotope}(\overline{\mathcal{Z}}_L,\underline{\mathcal{Z}}_L)\, , \quad 
(\overline{\mathcal{Z}}_L)_i = \argmaxB_j{(\overline{\mathcal{M}}_L)_{i,j}}\, ,  \quad
(\underline{\mathcal{Z}}_L)_i = \argminB_j{(\underline{\mathcal{M}}_L)_{i,j}}\, , \label{eq:zl} 
\end{align}
\begin{align}
[ \underline{\mathcal{M}}_L ,  \overline{\mathcal{M}}_L ]&= \textup{intervalMatrix}(\mathcal{M}_L),\label{eq:mlinterval}\\
\mathcal{M}_L &= X_{+} - \mathcal{M}_w - \tilde{M} \begin{bmatrix}1_{1 \times T}\\X_{-}-1 \otimes x^*\\ U_{-}-1 \otimes u^*\end{bmatrix},\label{eq:ml} \\
\mathcal{Z}_\epsilon &= \zono{0,\textup{diag}(L^\star \delta,\dots,L^\star \delta)}.
\end{align}
\end{theorem2}

\begin{proof}
We know from~\eqref{eq:fz_incl} that
\begin{align*}
    f(z) = (\tilde{M} + \Delta \tilde{M}) \begin{bmatrix} 1 \\ z - z^\star \end{bmatrix} + L(z),
\end{align*}
where $\Delta \tilde{M}$ captures the model mismatch defined by $\Delta \tilde{M} = \begin{bmatrix} f(z^\star) & \tilde{A} & \tilde{B} \end{bmatrix} - \tilde{M}$.
Hence, we need to show that $\mathcal{Z}_L + \mathcal{Z}_\epsilon$ over-approximates the modeling mismatch and the term $L(z)$, i.e.,
\begin{align*}
    \Delta \tilde{M} \begin{bmatrix} 1 \\ z - z^\star \end{bmatrix} + L(z) \in \mathcal{Z}_L + \mathcal{Z}_\epsilon
\end{align*}
for all $z \in \mathcal{F}$. For all $z_i \in \mathcal{D} = (U_-,X)$, we know that for some $(\hat{W}_-)_{\cdot, i} \in \mathcal{Z}_w$
\begin{align*}
    (X_+)_{\cdot,i} - (\hat{W}_-)_{\cdot, i} = (\tilde{M} + \Delta \tilde{M}) \begin{bmatrix} 1 \\ z_i - z^\star \end{bmatrix} + L(z_i),
\end{align*}
which implies
\begin{align}
    \Delta \tilde{M} \begin{bmatrix} 1 \\ z_i - z^\star \end{bmatrix} + L(z_i) \in (X_+)_{\cdot,i} - \mathcal{Z}_w - \tilde{M} \begin{bmatrix} 1 \\ z_i - z^\star \end{bmatrix}.
    \label{eq:deltamindata}
\end{align}
In this way, we can over-approximate the model mismatch and the nonlinearity term at one data point $z_i \in \mathcal{D} = (U_-,X)$. Extending right-hand side of \eqref{eq:deltamindata} to all the available data points in $\mathcal{D} = (U_-,X)$ and denoting the result by $\mathcal{M}_L$ yields \eqref{eq:ml}. We aim next to find one zonotope of the right-hand side of \eqref{eq:deltamindata} that is consistent with all the data points. This can be done by first converting $\mathcal{M}_L$ to the interval matrix in \eqref{eq:mlinterval}. Then we consider the lower and upper bound and convert the result to a zonotope $\mathcal{Z}_L$ in \eqref{eq:zl}.  
We can hence over-approximate the model mismatch and the nonlinearity term for all data points $z_i \in \mathcal{D} = (U_-,X)$, $i=0,1,\dots,T$, by
\begin{align*}
    f(z_i) \in \tilde{M} \begin{bmatrix} 1 \\ z_i - z^\star  \end{bmatrix} + \mathcal{Z}_L.
\end{align*}
Given the covering radius $\delta$ of our system together with Assumption~\ref{as:lipschitz}, we know that for every $z \in \mathcal{F}$, there exists a $z_i \in \mathcal{D} = (U_-,X)$ such that
    $\| f(z) - f(z_i) \| \leq L^\star \| z - z_i \| \leq L^\star \delta$.
This yields 
\begin{align*}
    f(z) \in \tilde{M} \begin{bmatrix} 1 \\ z_i - z^\star \end{bmatrix} +  \mathcal{Z}_L + \mathcal{Z}_\epsilon,
\end{align*}
with $\mathcal{Z}_\epsilon = \zono{0,\text{diag}(L^\star \delta,\dots,L^\star \delta)}$.
\end{proof}
For an infinite amount of data, i.e., $\delta \rightarrow 0$, we can see that $\mathcal{Z}_\epsilon \rightarrow 0$, i.e., the formal $\mathcal{Z}_L$ then fully captures the modeling mismatch and the Lagrange reminder. While $\mathcal{Z}_{\epsilon}$ is needed for guarantees, still neglecting this term provides over-approximations of the reachable set in the numerical examples given sufficient data.
\vspace{-2mm}
\begin{remark}
\label{rem:approx}
Note that determining $L^\star$ as well as computing $\delta$ is non-trivial in practice. If we assume that the data is evenly spread out in the compact input set of $f$, then the following can be a good approximation of the upper bound on $L^\star$ and $\delta$:
\begin{align*}
    \hat{L}^\star = \max_{z_i, z_j \in D, i\neq j}  \frac{\|f(z_i) - f(z_j)\|}{\| z_i - z_j \|}, \quad \quad
    \hat{\delta} = \max_{z_i \in D} \min_{z_j \in D, j \neq i} \| z_i - z_j \|.
\end{align*}
Other methods to calculate the Lipschitz constant $L^\star$ can be found in~\citep{conf:montenbruckLipschitz,conf:novaraLipschitz}, and a sampling strategy to obtain a specific $\delta$ is introduced in~\citep{conf:montenbruckLipschitz}.
\end{remark}
\vspace{-5mm}
%
\begin{remark}
We choose the linearization points as the center of the current input zonotope $\mathcal{U}_k$ and state zonotope $\hat{\mathcal{R}}_k$, and we repeat the linearization at each time step $k$. In model-based reachability analysis, the optimal linearization point is the center of the current state and input zonotopes as proved in \cite[Corollary 3.2]{conf:thesisalthoff}, which then minimizes the set of Lagrange remainders. Therefore, choosing the center of the current input and state zonotopes as linearization points is a natural choice, while the theoretical results are independent of this choice.
\end{remark}
\vspace{-5mm}

\vspace{-1mm}
\section{Evaluation}\label{sec:eval}
\vspace{-1mm}

\begin{figure}[!tbp]
    \begin{tabular}{ p{0.320\textwidth}  p{0.320\textwidth} p{0.320\textwidth}}
        \includegraphics[scale=0.25]{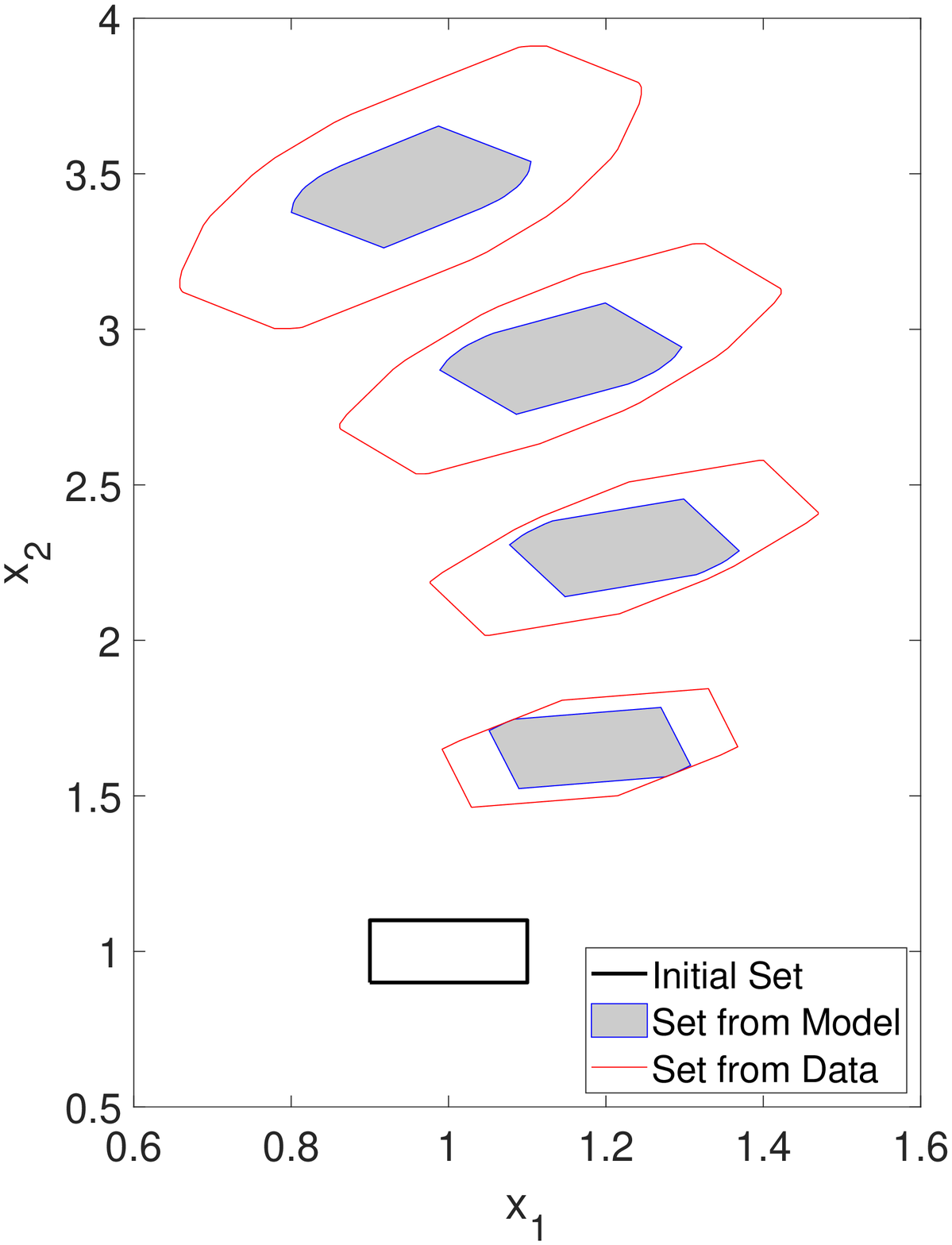}
   &
        \includegraphics[scale=0.25]{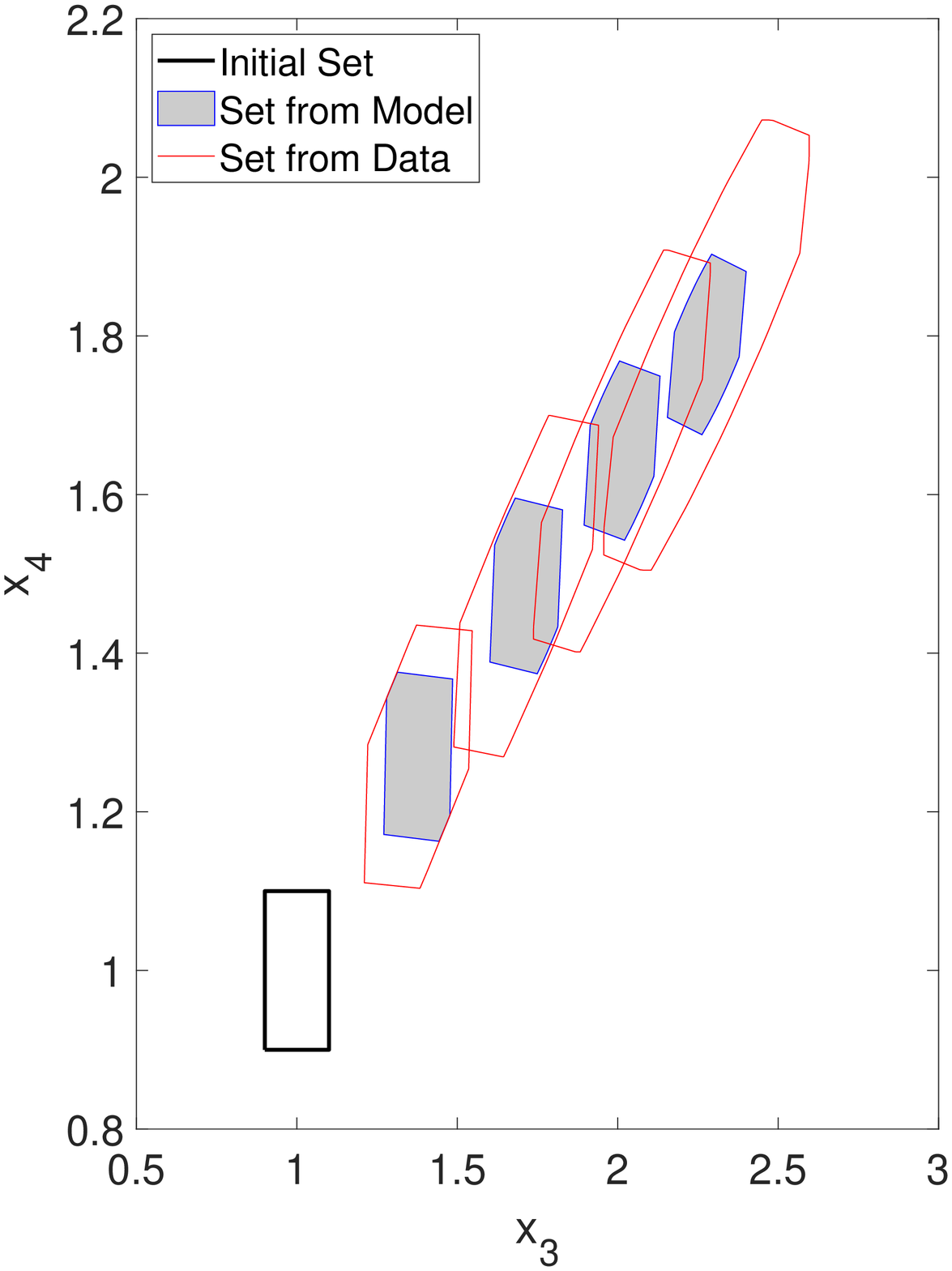}
         &
        \includegraphics[scale=0.25]{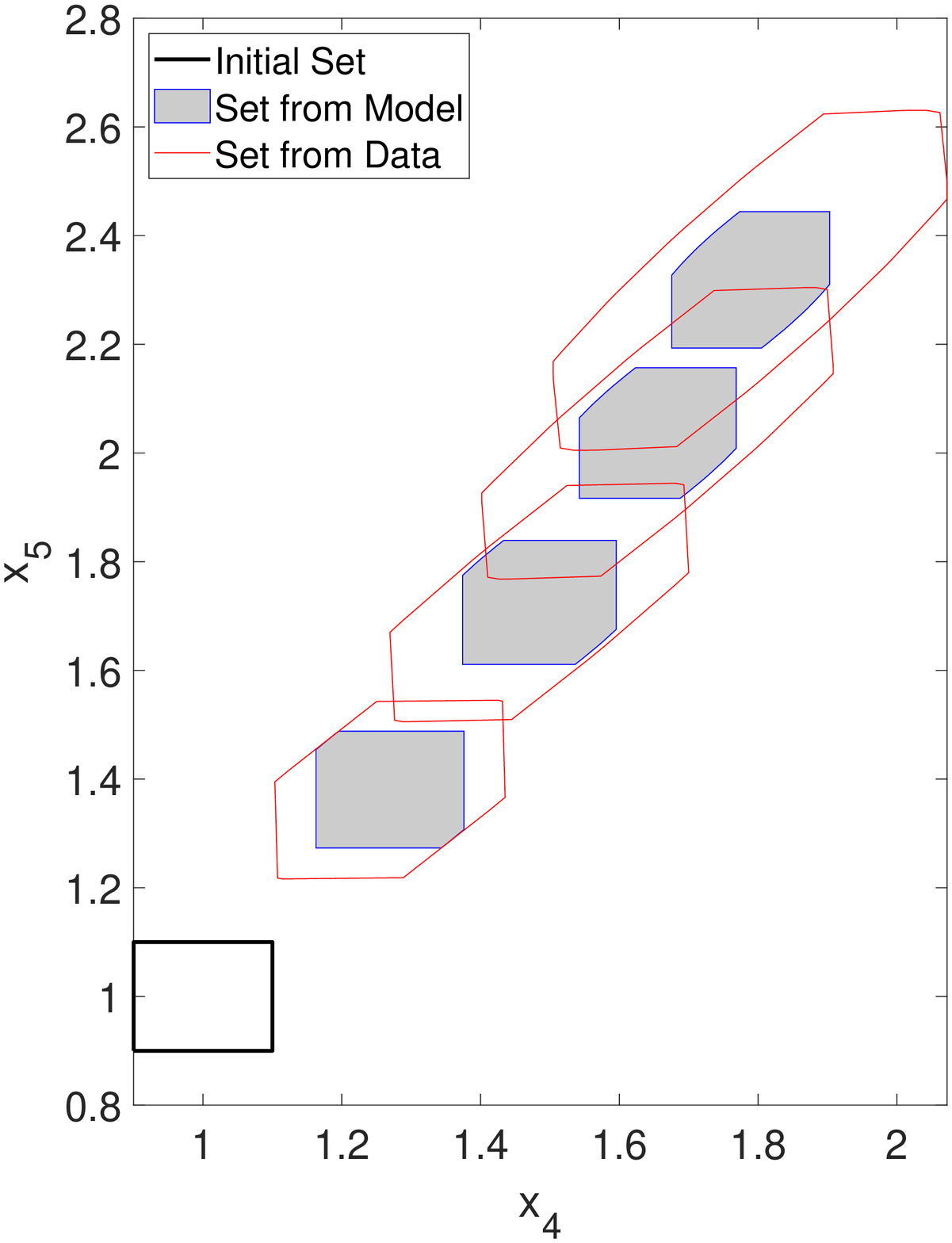}
  \end{tabular}
  \vspace{-6mm}
\caption{The reachable sets of the system in \eqref{eq:sysexamplediscrete} computed via Theorem \ref{th:reach_lin} from noisy data.}
    \label{fig:reach}
      \vspace{-3mm}
\end{figure}

To demonstrate the usefulness of the presented approach, we consider the reachability analysis of a five dimensional system which is a discretization of the system used in \citep[p.39]{conf:thesisalthoff} with sampling time $0.05$ sec. The system has the following parameters

\begin{align}
\begin{split}
A&=\begin{bmatrix}
    0.9323 &  -0.1890   &      0   &      0   &      0 \\
    0.1890 &   0.9323  &       0  &       0   &      0 \\
         0 &        0  &  0.8596  &   0.0430  &        0 \\
         0 &         0   & -0.0430    & 0.8596      &    0 \\
         0 &         0  &        0    &      0   &  0.9048
\end{bmatrix}, \\
B&=\begin{bmatrix} 
    0.0436&
    0.0533&
    0.0475&
    0.0453&
    0.0476
    \end{bmatrix}^T.
\end{split}
\label{eq:sysexamplediscrete}
\end{align}
The initial set is chosen to be $\mathcal{X}_0=\zono{1,0.1 I}$ where $1$ and $I$ are vectors of one and the identity matrix, respectively. The input set is $\mathcal{U}_k=\zono{10,0.25}$. We consider computing the reachable set when there is random noise sampled from the zonotope  $\mathcal{Z}_w=\zono{0,\begin{bmatrix}0.005& \dots & 0.005\end{bmatrix}^T}$. We make use of $65$ data points. The zonotopic reachable sets using the model and using the introduced approach in Theorem \ref{th:reach_lin} are shown in Figure \ref{fig:reach}. 
The presented approach guarantees an over-approximation of all the reachable set of all models with the data, hence over-approximating the reachability set of the system in \eqref{eq:sysexamplediscrete}. 

\begin{figure}[t!]
    \centering
    \includegraphics[scale=0.4]{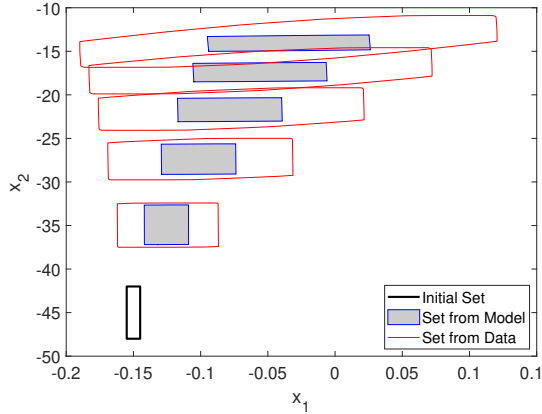}
    \vspace{-6mm}
    \caption{The data-driven and model-based reachable sets of the nonlinear system.}
    \label{fig:reachnonlinear}
    \vspace{-5mm}
\end{figure}


The proposed nonlinear data-driven reachability analysis is applied to a nonlinear continuous stirred tank reactor (CSTR) simulation model \citep{conf:nonlinearexample}. The initial set is a zonotope $\mathcal{X}_0 =\zono{\begin{bmatrix} 0.015 \\ -45 \end{bmatrix},\begin{bmatrix} 0.005 & 0 \\ 0 & 3 \end{bmatrix}}$. The input set $\mathcal{U}_k =\zono{1,\begin{bmatrix} 0.1 & 0 \\ 0 & 2 \end{bmatrix}}$ and the noise set $\mathcal{Z}_w=\zono{0,\begin{bmatrix}0.01 & 0.01\end{bmatrix}^T}$. We apply the results from Theorem \ref{th:reachdisnonlin} and neglect the term $\mathcal{Z}_{\epsilon}$. The result is plotted in Figure \ref{fig:reachnonlinear} and shows that the model-based reachable region is correctly over-approximated. 

\vspace{-3mm}
\section{Conclusion}\label{sec:con}
\vspace{-2mm}
We considered the problem of computing the reachable regions directly from noisy data without a priori model information.
Assuming knowledge of a bound on the noise in the data, we first provided a computationally simple approach to guarantee over-approximation of the reachable set of an unknown linear system by over-approximating the reachable set of all systems consistent with the data and the noise bound. 
Moreover, we consider Lipschitz nonlinear systems, where we first fitted a linear model and then over-approximated the model mismatch and Lagrange reminder from data, resulting again in a guaranteed over-approximation of the reachable set. 
\vspace{-3mm}
\section*{Acknowledgement}
\vspace{-2mm}
This work was supported by the Swedish Research Council, the Knut and Alice Wallenberg Foundation, as well as the Deutsche Forschungsgemeinschaft (DFG, German Research Foundation) under Germany’s Excellence Strategy - EXC 2075 - 390740016. The authors thank the International Max Planck Research School for Intelligent Systems (IMPRS-IS) for its support.
%



\bibliography{ref} 

\end{document}